# Toward a Comprehensive Model of Snow Crystal Growth:
# 8. Characterizing Structure-Dependent Attachment Kinetics near -14 C

Kenneth G. Libbrecht

Department of Physics
California Institute of Technology
Pasadena, California 91125
kgl@caltech.edu

**Abstract.** In this paper I examine snow crystal growth near -14 C in comparison with a comprehensive model that includes Structure-Dependent Attachment Kinetics (SDAK). Analyzing a series of ice-growth observations in air, I show that the data strongly support the model, which stipulates that basal growth is described by classical terrace nucleation on faceted surfaces in this temperature region. In contrast, prism growth exhibits a pronounced "SDAK dip" that substantially reduces the nucleation barrier on narrow prism facets (relative to that found on broad prism facets). I use these measurements to further characterize and refine the SDAK model, which effectively explains the robust formation of platelike snow crystals in air near -14 C.

## ❆ Introduction

As suggested by its title, this series of papers represents my ongoing efforts to develop a comprehensive model of the physical dynamics of snow crystal growth. The research remains a work in progress, as the formation of ice crystal structures from the solidification of water vapor is a remarkably rich and subtle phenomenon, involving the complex interplay of many physical processes. Some past reviews of this subject can be found in [1954Nak, 1987Kob, 2005Lib, 2017Lib], and a new book on the subject is currently in press [2021Lib].

A particularly intriguing aspect of snow crystal growth is illustrated in the Nakaya diagram, which summarizes ice growth morphologies in air as a function of temperature and water-vapor supersaturation [1954Nak]. The chart has appeared in numerous forms over many decades [1958Hal, 1961Kob, 1990Yok, 2009Bai, 2012Bai, 2017Lib, 2021Lib], and one variation is shown in Figure 1.

The well-known appearance of platelike snow-crystal forms near -2 C, columnar forms near -5 C, especially thin platelike forms near -14 C, and columnar forms again around -40 C has been an enduring scientific puzzle for nearly 75 years. Models based on surface diffusion [1958Hal, 1963Mas] and temperature-dependent surface roughening [1982Kur, 1984Kur] have been proposed, but these models are generally not well supported by modern quantitative growth measurements [2021Lib]. I recently proposed a new Comprehensive Attachment Kinetics (CAK) model that may finally provide a physical model that explains and quantifies many essential aspects of the Nakaya diagram [2019Lib1, 2021Lib].

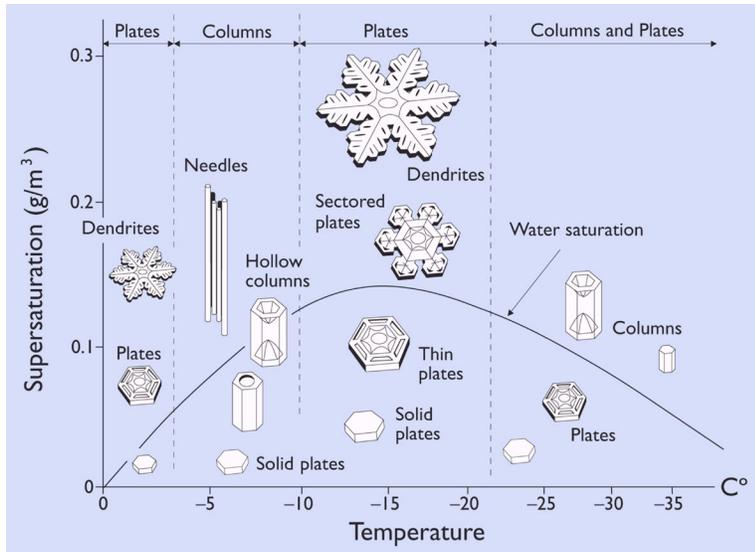

Figure 1: The *Nakaya diagram* illustrates snow crystal morphology as a function of temperature and supersaturation, where the latter is shown here as the "excess" water vapor density in the air, above the value for saturated air. The *water saturation* line further shows the supersaturation of liquid water with respect to ice. Explaining why snow crystals exhibit such varied growth behaviors at different temperatures and supersaturations is a remarkably challenging scientific puzzle, with many aspects still unexplained.

Regarding the temperature-dependent growth behavior seen in the Nakaya diagram, there appears to be a reasonable consensus on two general statements: 1) the temperature-dependent growth behavior is largely the result of temperature-dependent attachment kinetics at the ice/vapor interface [1984Kur], and 2) terrace nucleation plays a major role in the attachment kinetics, bringing about the appearance of strong basal and prism facets [1998Nel]. In addition, it also appears that the attachment kinetics on large basal and prism faceted surfaces is sometimes quite different from that on narrow facets. On the latter, edge effects can rather strongly reduce the effective nucleation barrier compared to broad facets, a phenomenon I have called Structure Dependent Attachment Kinetics (SDAK) [2003Lib1].

The CAK model provides an overarching physical picture of the attachment coefficients $\alpha_{basal}$ and $\alpha_{prism}$ as a function of supersaturation and temperature from approximately -1 C to -30 C, which are strongly influenced by the SDAK effect. This semi-empirical model appears to reproduce most of the salient features of the snow-crystal Nakaya diagram over this substantial temperature range, with predicted growth behaviors that have been confirmed by subsequent measurements [2019Lib2, 2020Lib]. The model is based on plausible assumptions regarding the molecular dynamics of terrace nucleation, surface diffusion, and other factors, and it nicely explains how both platelike and columnar morphologies can be readily observed under different conditions at -5 C [2012Kni, 2019Lib2], indicating a subtle sensitivity to initial conditions.

In the present paper, I continue developing the CAK model by examining ice-growth data near -14 C as a function of supersaturation and temperature. As described below, the CAK model predicts a pronounced reduction in the prism nucleation barrier near -14 C from the SDAK mechanism, which can be quantified using a series of ice-growth measurements. The data reveal a remarkably good agreement between model predictions and measured parameters, strongly supporting the importance of the SDAK phenomenon in overall terms. The data also provide an accurate parameterization of the basal and prism attachment kinetics as a function of temperature and surface supersaturation near -14 C, facilitating realistic numerical modeling of snow crystal growth dynamics and many additional experimental investigations of this fascinating phenomenon.

# ❋ The Comprehensive Attachment Kinetics (CAK) Model

As described in [2021Lib], the Comprehensive Attachment Kinetics (CAK) model begins by characterizing the nucleation-limited growth of large basal and prism facets. Because much of the needed background information for this aspect of the CAK model has been provided in prior publications [2021Lib, 2019Lib1, 2019Lib2], I summarize briefly this aspect of the basal and prism attachment kinetics here, focusing on temperatures near -14 C.

## Large Faceted Surfaces

On faceted ice/vapor surfaces that are large enough to be unaffected by edge effects, the CAK model provides that the attachment coefficients are given by

$$\alpha_x(\sigma_{surf}) = A_x e^{-\sigma_{0,x}/\sigma_{surf}} \quad (1)$$

where $x$ stands for either *basal* or *prism*, $\sigma_{surf}$ is the water-vapor supersaturation at the ice/vapor interface, $\sigma_{0,x}$ is a nucleation parameter that derives from the terrace step energy at the ice/vapor interface, and $A_x$ depends on the normal admolecule surface diffusion and other parameters. This functional form is provided by classical terrace nucleation theory [e.g., 1994Ven, 1996Sai, 1999Pim, 2002Mut], but the parameters relating to ice growth from water vapor must be determined from experimental measurements.

Figure 2 shows the empirical CAK model parameters as a function of temperature for large basal and prism facets. Theory tells us that the $\sigma_{0,x}$ parameter derives mainly from the terrace step energy on faceted surfaces, and molecular-dynamics simulations can be used to calculate this quantity [2020Llo]. The calculations are still of rather low accuracy, however, so experimental measurements provide the estimates shown in Figure 2 [2021Lib, 2020Lib, 2013Lib]. The $A_x$ parameter is quite difficult to ascertain from theory, so again the values in Figure 2 derive from experiments. In both cases, smooth curves were drawn as approximate fits to the data, and these curves specify the CAK model. The different empirical curves plotted in Figure 2 are

$$\sigma_{0,basal}(T_*) = 0.02 T_*^{1.75} + 0.3 \quad (2)$$

$$\sigma_{0,prism}(T_*) = 0.02 T_*^{1.9} - 0.025(T_* - 0.3) \quad (3)$$

$$A_{basal} = 1 \quad (4)$$

$$A_{prism} = (0.4 + 0.04|T_* - 4|^3) / (2.2 + 0.04|T_* - 4|^3) \quad (5)$$

where $T_* = (T_m - T)$ is measured in Celsius and $\sigma_{0,x}$ is measured in percent. These functional forms were chosen solely to provide reasonable fits to the available data. An in-depth discussion of the experiments and data analysis techniques leading to these curves can be found in [2021Lib]. Once again, these results apply only to faceted surfaces that are effectively of infinite extent, so they are not affected by edge effects implicit in the SDAK mechanism described below. Moreover, the model assumes that the parameters in Figure 2 are independent of background air pressure. As the present paper examines only temperatures up to -10 C, we can assume $A_{prism} \approx 1$ in this regime.

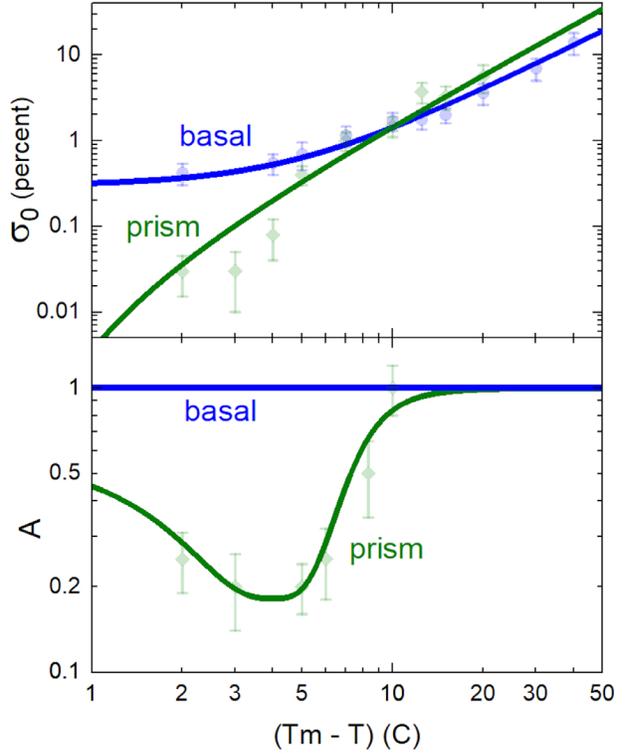

Figure 2: The graphs above define the CAK model of the attachment kinetics on large basal and prism surfaces, drawn as smooth curves for clarity [2021Lib]. As described in the text, the model parameters assume a terrace-nucleation model, as this is well supported by growth measurements [2013Lib]. The present paper focuses on temperatures between -10 C and -30 C, where $A_{basal} \approx A_{prism} \approx 1$.

### STRUCTURE-DEPENDENT ATTACHMENT KINETICS (SDAK)

The phenomenon of Structure-Dependent Attachment Kinetics has also been described in previous publications [2021Lib, 2019Lib1, 2003Lib1], but I outline its salient features here, as the present paper continues the development our physical model and its effects on snow crystal growth.

Figure 3 outlines the basic mechanism responsible for the SDAK effect (according to the CAK model), which is described in greater detail in [2021Lib, 2019Lib1]. Near -14 C, the model depends on surface diffusion that transports water molecules from the sharpened basal/prism corners to the uppermost prism terrace on the edge of a thin plate. At temperatures far below -14 C, this surface diffusion is strongly inhibited by the Ehrlich-Schwoebel (E-S) barrier, which is commonly discussed in crystal-growth textbooks [1996Sai, 1999Pim, 2002Mut]. In this low-temperature circumstance, increasing the applied supersaturation sharpens the corners as shown in Figure 3b and there is little change in the attachment kinetics, which remains essentially the same as the large-facet kinetics described in Figure 2.

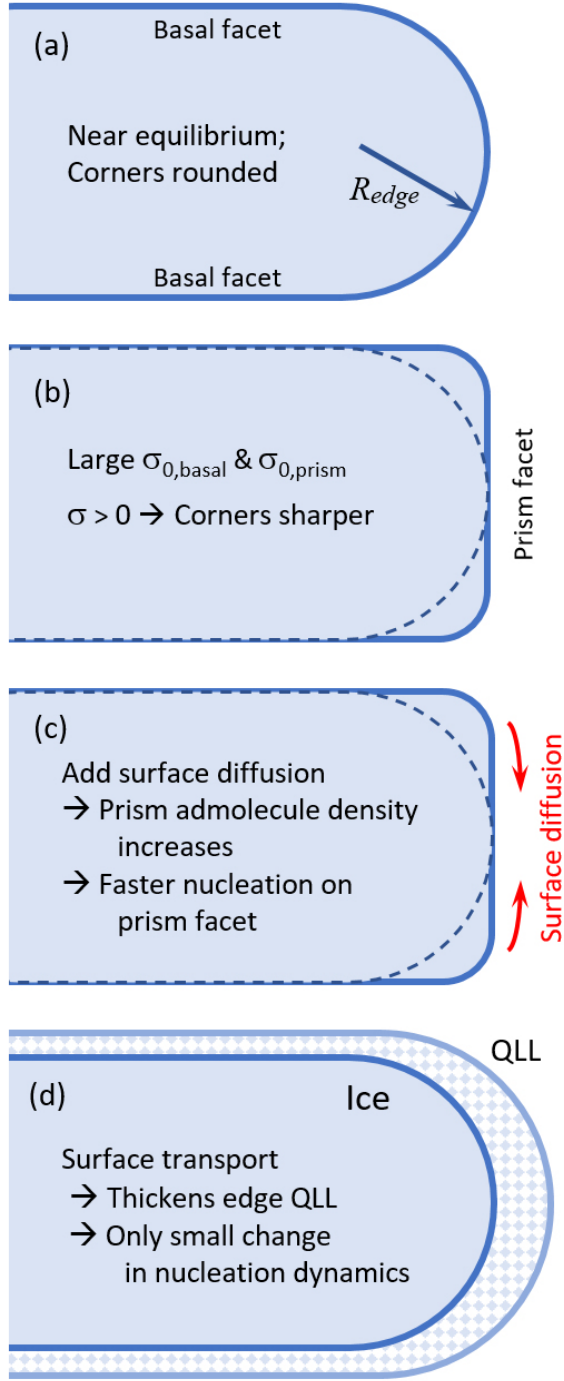

Figure 3: A schematic explanation of the phenomenon of Structure-Dependent Attachment Kinetics (SDAK). (a) When the applied supersaturation is low, the edge of a thin plate is rounded, as this shape minimizes the surface energy. (b) At higher supersaturation, the corners become sharpened by the Gibbs-Thomson effect while the basal and prism growth is inhibited by a strong terrace nucleation barrier. (c) If surface diffusion is significant onto the prism facet, the admolecule density will increase, yielding an increased rate of nucleation on the prism surface. (d) If strong surface premelting is present, surface diffusion increases the QLL thickness on the prism facet, but this does not increase the growth rate appreciably, as nucleation now occurs at the ice/QLL interface [2021Lib, 2019Lib1].

As the temperature approaches -14 C, however, the CAK model postulates that surface premelting on the prism facet (but not on the basal facet) begins to break down the E-S barrier, thus allowing some surface diffusion. Because the sharpened corners present a non-equilibrium shape with increased surface energy, there is a thermodynamic potential that drives surface diffusion as shown, as this transport tends to round the edge and thereby lower its overall surface energy. Put another way, surface tension provides a thermodynamic force that drives corner-to-facet surface diffusion, as this material transport will reduce the overall surface energy.

As shown in Figure 3c, the resulting surface diffusion brings additional admolecules to the upper prism terrace, thus increasing the admolecule density above its normal level. Because the nucleation probability increases with increasing admolecule density, the additional surface transport effectively reduces the nucleation barrier on the uppermost prism terrace [2021Lib, 2019Lib1]. According to the CAK model, this SDAK effects is most prevalent on the prism facet when the temperature is near -14 C, resulting from the postulated onset of surface premelting at that temperature. Note that terrace nucleation is still necessary for growth on the top prism terrace; the additional admolecules simply facilitate the nucleation process, effectively lowering the nucleation barrier.

At temperatures above -14 C, surface premelting yields even greater surface diffusion, but it also increases the overall thickness of the quasi-liquid layer (QLL) on the ice surface. When the QLL is sufficiently thick, terrace nucleation effectively takes place at the QLL/ice interface, where the concept of an admolecule surface density is no longer relevant. In this regime, the QLL/ice interface begins to act essentially like a water/ice interface, so terrace nucleation is not much affected by additional surface diffusion, as illustrated in Figure 3d.

Putting all the pieces together, this rough physical model of the SDAK phenomenon predicts an "SDAK dip" in the effective prism nucleation barrier, as illustrated in Figure 4. As postulated in the CAK model, the dip in $\sigma_{0,prism}$ near -14 C is accompanied by a similar dip in $\sigma_{0,basal}$ near -5 C. As this point the CAK model is far beyond what any molecular theory could meaningfully predict, as this would require a detailed understanding of how surface premelting affects the E-S barrier as a function of temperature on the basal and prism facets. With little guidance from molecular theory, the positions and widths of the SDAK dips shown in Figure 4 were judiciously chosen to provide a reasonable explanation for the temperature-dependent behaviors seen in the Nakaya diagram [2021Lib, 2019Lib1].

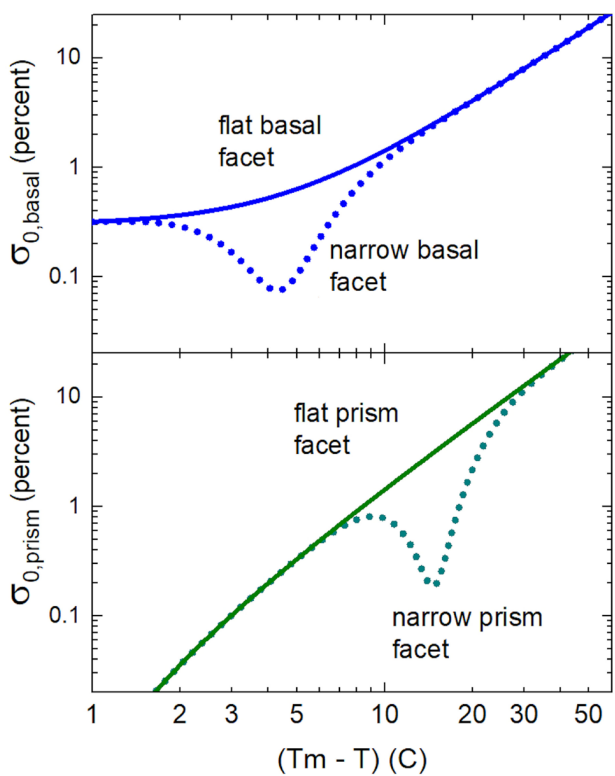

Figure 4: The SDAK phenomenon shown in Figure 3 leads to localized "SDAK dips" in the effective terrace nucleation barriers on narrow basal and prism facets [2021Lib, 2019Lib1]. The CAK model assumes slight differences in the SDAK phenomenon on the basal and prism facets, thus yielding the SDAK dips shown here.

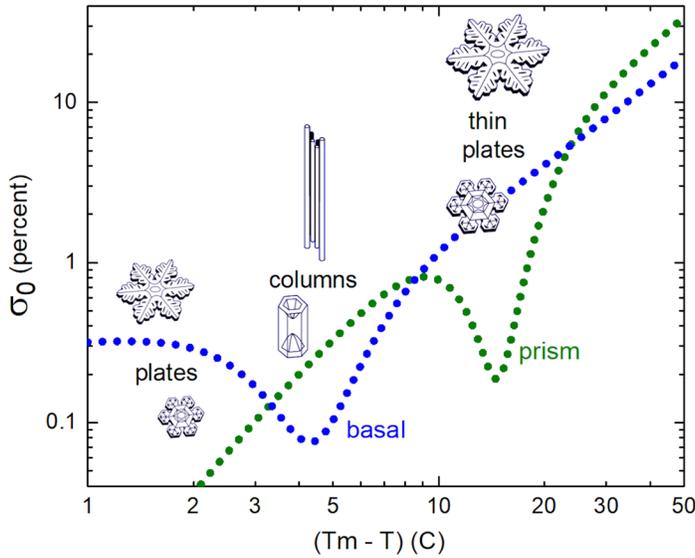

Figure 5: The temperature-dependent changes in the effective nucleation parameters show how the CAK model can provide a ready explanation for much of the temperature-dependent behavior observed in the Nakaya diagram, specifically the different transitions between platelike and columnar growth [2021Lib].

Plotting the two SDAK dips in Figure 4 on a single graph yields the result shown in Figure 5. Because a lower nucleation barrier means faster growth, we see that the CAK model explains the prevalence of platelike forms near -2 C, needle-like crystals near -5 C, especially thin plates near -14 C, and columnar forms around -40 C, reproducing the known growth behaviors in the Nakaya diagram. While the CAK model, including the SDAK mechanism, contains many speculative features, it appears to be (in my opinion) the only existing model that can reasonably explain this aspect of snow crystal growth while remaining consistent with a substantial body of ice-growth measurements.

Figure 6: The SDAK phenomenon can be divided into separate cases. In the two-sided case (a), surface diffusion from two basal/prism corners deposits admolecules onto a narrow prism facet, producing a large nucleation enhancement. In the one-sided case (b), surface diffusion from a single basal/prism corner deposits admolecules onto the adjacent prism facet, leading to a lesser nucleation enhancement on the top prism terrace of an overall sloped vicinal surface.

Considering the SDAK mechanism in more detail (which will become relevant below when examining growth data), we can anticipate the two different cases illustrated in Figure 6. The two-sided case was described in detail in Figure 3, as this produces the greatest enhancement in admolecule density and nucleation dynamics. This case applies to thin platelike crystals, where the width of the top prism terrace is $w \approx \sqrt{8aR_{edge}}$, where $a$ is the molecule size and $R_{edge}$ is the radius of curvature of the plate edge shown in Figure 3. For a typical thin-plate snow crystal growing near -14 C, we might have $R_{edge} \approx 1$ μm and $w \approx 50$ nm.

With such a thin plate, surface diffusion from both nearby basal/prism corners deposits admolecules on the top prism terrace, as shown in Figure 6a. Because the top terrace is so narrow, this can lead to a large increase in admolecule surface density, determined by a balance of: 1) deposition from the vapor phase, 2) sublimation back into the vapor phase, and 3) surface diffusion from the two basal/prism corners.

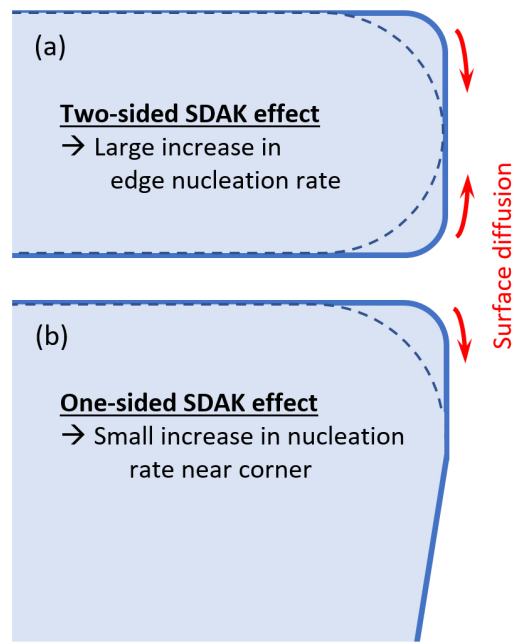

The admolecule dynamics is somewhat different in the one-sided case, as illustrated in Figure 6b. Now the admolecule surface density on the top prism terrace is determined by four processes: 1) deposition from the vapor phase, 2) sublimation back into the vapor phase, 3) surface diffusion from the basal/prism corner, and 4) surface diffusion from the top terrace onto the neighboring vicinal surface. The SDAK mechanism again leads to an increase in the admolecule surface density near the corner region, because the fourth process above goes to zero if there is no such increase. However, one expects that the SDAK phenomenon will be weaker overall in the one-sided case when compared with the two-sided case.

Note that the fourth process in the above list is also inhibited by the E-S barrier, and surface-tension will not strongly drive surface diffusion over this barrier (like it does with the diffusion from the corners). Moreover, the spacing between adjacent terrace steps can be quite small on a vicinal surface, of order 20 nm with just a one-degree vicinal angle. Thus, we expect that surface diffusion onto the top prism terrace (from the basal/prism corner) will be quite strongly driven compared with surface diffusion off the top terrace (onto the neighboring vicinal surface). By this logic, the one-sided SDAK effect may still be quite significant, although never as strong as the two-sided version.

One undeniable and favorable aspect of the CAK model, which incorporates the SDAK effect, is that it makes numerous predictions that can be tested using targeted experimental investigations. The overarching goal of this paper, therefore, is to treat the CAK model as a scientific hypothesis in need of testing. Our next step, therefore, is to describe a series of ice-growth experiments designed to test the model as thoroughly as possible.

## ❄ Ice Growth on Electric Needles

I have found that ice growth on "electric" ice needles, as illustrated in Figure 7, can be especially useful for characterizing the SDAK effect. To create the set of crystals in this figure, first a thin wire was placed inside a vertical diffusion chamber with the wire tip near -6 C in highly supersaturated air, with the local supersaturation being somewhat above 100%. Applying 2000 volts to the wire stimulated the growth of thin ice needles, and a slight additional vapor of acetic acid caused the e-needles to grow out along the ice c-axis. Detailed descriptions of the apparatus and e-needle physics can be found in [2021Lib, 2014Lib1].

After the e-needles grew out to lengths of 2-3 mm, the wire was moved to an adjoining second vertical diffusion chamber, where the temperature and supersaturation could be controlled separately. In the example shown in

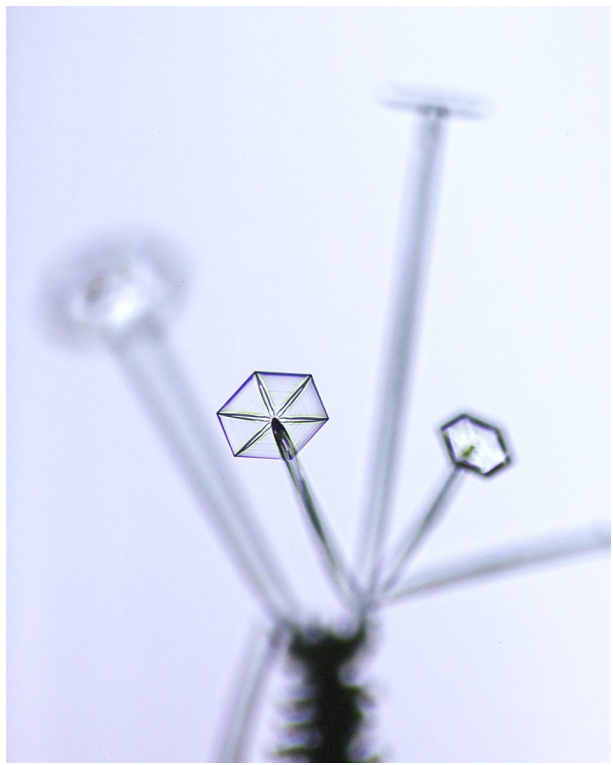

Figure 7: Platelike snow crystals grow on the ends of "electric" ice needles at a temperature near -15 C in filtered laboratory air at a pressure one bar [2021Lib]. The e-needles shown here are about 2-3 mm in overall length.

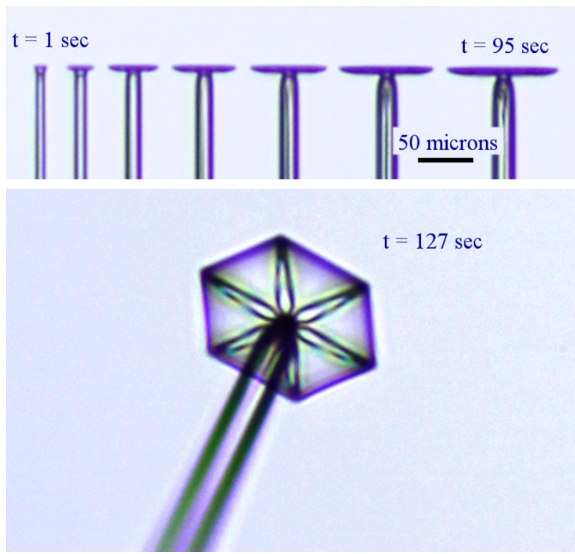

Figure 8: A typical e-needle growth run showing the formation of a platelike crystal on an ice needle as a function of time. From the side-view images (top series), both the basal and prism growth rates could be measured. The bottom image is the same crystal from a different viewing angle, showing the hexagonal-plate morphology with six ridges on the lower basal surface [2021Lib].

Figure 7, the chamber center temperature (where the needles were located) was set to -15 C with a local supersaturation of approximately 16%. In these conditions, thin hexagonal plates grew on the tips of the e-needles. The air in the second diffusion chamber was also continuously cleaned via charcoal filtration to reduce any deleterious effects from unwanted chemical contaminants.

Note that the supersaturation reported here (and in Figure 1) gives the supersaturation in the air when no growing crystals are present, which I call the "far-away" supersaturation $\sigma_{far}$ (also often called $\sigma_\infty$). Because ice growth in air is strongly limited by particle diffusion, $\sigma_{far}$ is generally much larger than $\sigma_{surf}$, which is the supersaturation at the ice/vapor interface.

Using this apparatus, I grew many similar e-needle clusters at temperatures ranging from -8 C to -30 C with supersaturations ranging from 2% to 128%, usually in steps of $\sqrt{2}$. The temperatures were measured by swinging a small thermistor into the test region, and the supersaturation was determined by modeling the linear diffusion chamber [2021Lib, 2014Lib1]. In a typical run, the e-needle with the highest vertical reach was selected by orienting it perpendicular to the image plane, giving a side view of the resulting growth, as illustrated in Figure 8. From such a series of images, the overall growth of the crystal, including the basal and prism growth velocities of the platelike crystal, could be determined.

Figure 9 shows a comparison of the growth data in Figure 8 with several computational models of diffusion-limited growth

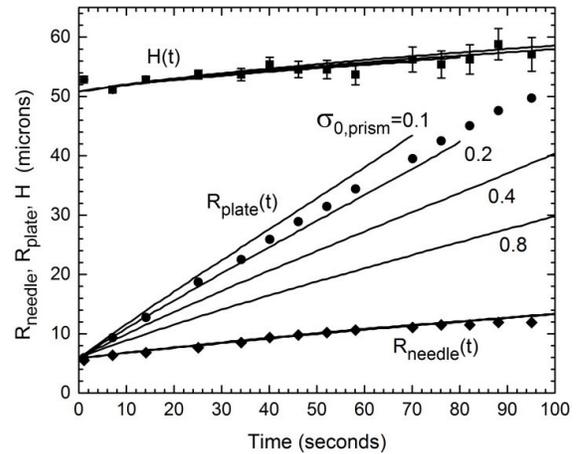

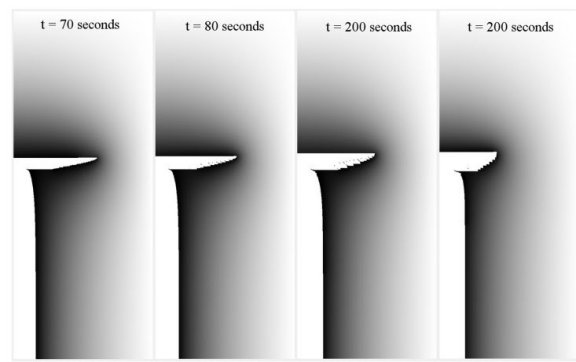

Figure 9: Measurements of the crystal growth as a function of time for the example shown in Figure 8 [2015Lib2]. In addition to the measured data points, several lines show models of the growth using different parameterizations of the attachment kinetics. The lower panels show cross sections of the computational models, where brightness is proportional to supersaturation. Additional details can be found in [2015Lib2].

[2015Lib2]. In this "forward modeling" analysis, the attachment coefficients $\alpha_{basal}(\sigma_{surf})$ and $\alpha_{prism}(\sigma_{surf})$ were specified in the model assuming a known physical parameterization, and corresponding model crystals could then be compared with experimental measurements. This analysis strategy is quite time-consuming, as many models must be calculated and compared with the data, and the computational model itself includes numerous implicit assumptions regarding the numerical techniques used for calculating the diffusion-limited growth of faceted crystals.

In the present paper, I performed a substantially simpler "witness surface" analysis that did not require any detailed diffusion modeling. In this analysis, the basal growth was assumed to be consistent with the large-facet CAK model shown in Figure 2. With this assumption, the basal growth rate could be used to directly determine the maximum surface supersaturation along the faceted surface, as this value of $\sigma_{surf}$ controlled the overall basal growth rate. Assuming the CAK model is correct for faceted basal growth below -10 C, the top basal terrace becomes a "witness" surface, in that its growth velocity can be used to determine $\sigma_{surf}$ using $v_{basal} = \alpha_{basal}(\sigma_{surf})v_{kin}\sigma_{surf}$.

Using the witness-surface method to ascertain $\sigma_{surf}$ at some chosen time during the growth series (from the measured $v_{basal}$ at that time), I then assumed that $\sigma_{surf}$ was essentially identical on nearby basal and prism surfaces. Then the measured $v_{prism}$ gives $\alpha_{prism}$ from $v_{prism} = \alpha_{prism}(\sigma_{surf})v_{kin}\sigma_{surf}$. This analysis can be applied to complex crystal morphologies because the growth is largely "facet dominated" [2021Lib], meaning that the overall development of the crystal is determined mainly by the growth rates of the slowest-growing surfaces, which are typically the uppermost basal and prism facets.

The key element in this witness-surface analysis is the assumption that $\alpha_{basal}(\sigma_{surf})$ is completely known from the CAK model, given by terrace nucleation on large basal facets with the parameterization quantified in Figure 2. If this assumption is correct (and I believe it is, to reasonable accuracy, based on experimental data leading to the development of the CAK model), then the experiments described below allow us to use $\alpha_{basal}$ to determine $\sigma_{surf}$ and then $\alpha_{prism}$. In this way, the witness-surface analysis provides a "bootstrap" process that allows us to use our understanding of $\alpha_{basal}$ to measure the less-well-known $\alpha_{prism}$ and thus better characterize the SDAK phenomenon on narrow prism facets.

A lesser assumption in the witness-surface analysis is that $\sigma_{surf}$ is essentially identical on nearby basal and prism surfaces. This supposition is expected to be reasonably accurate on the tips of dendritic crystals, as the uppermost basal and prism terraces are quite close to one another in such sharp-tipped morphologies. The assumption is less accurate on crystals with broad facets, and it can be quite poor with long columnar crystals (for example, see [2020Lib]). Systematic errors of perhaps a factor of two in $\sigma_{surf}$ may be present using the witness-surface analysis technique for some crystal morphologies, but we will see below that such uncertainties will not appreciably alter the overall conclusions reached in this paper.

## Results at -14 C

Beginning with the highest supersaturation level at $\sigma_{far} = 128\%$, Figure 10 shows a typical growing crystal. Although the six dendritic branches have complex morphological structures with substantial sidebranching (not seen in this side-view figure), each branch tip must include a single small basal facet with an uppermost basal terrace, along with two nearby small prism facets. The value of $\sigma_{surf}$ on these surfaces is determined largely by particle diffusion surrounding the entire structure, and the facet growth rates are limited by terrace nucleation on the top terraces. Following the CAK model at -14 C, the basal growth rate is assumed to be given by $v_{basal} =$ $\alpha_{basal} v_{kin} \sigma_{surf}$ with $\alpha_{basal} = \exp(-\sigma_{0,basal}/\sigma_{surf})$ and $\sigma_{0,basal} = 2.33$ percent at -14 C. From this, $\sigma_{surf}$ can be determined from the measured $v_{basal}$, and I further assume that this value of $\sigma_{surf}$ applies to both the basal and prism facets at the branch tips. Then $\alpha_{prism}$ can be derived from the measured $v_{prism}$ using $v_{prism} = \alpha_{prism} v_{kin} \sigma_{surf}$.

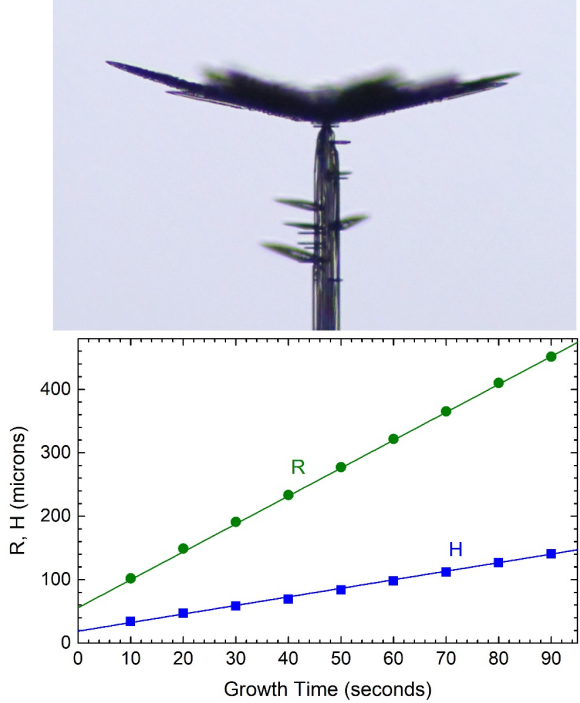

Figure 10: (Top) A side view of a snow crystal growing on the end of an e-needle at -14 C in normal air at a supersaturation level of 128%. The crystal is essentially a fernlike stellar dendrite with a slightly conical overall shape. The ratio of prism to basal growth rates at the dendrite tips is about 3.5:1. (Bottom) Measurements of the tip radius R and tip height H as a function of time for this crystal (with H measured relative to some arbitrary fixed point). The basal and prism growth velocities of the tip equal dH/dt and dR/dt, respectively. The image is from the last point of the time series.

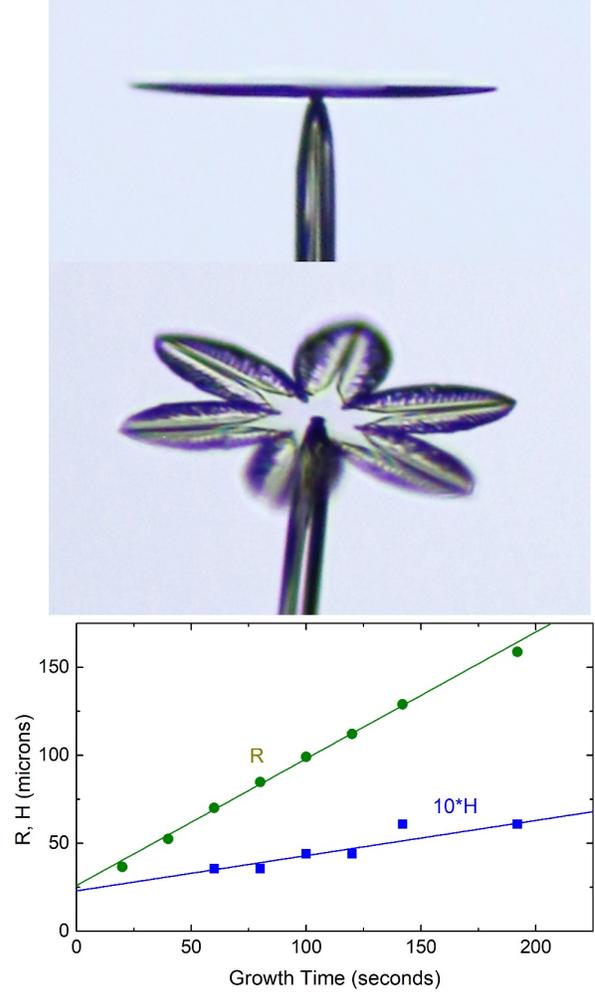

Figure 11: Two views of a snow crystal growing on the end of an e-needle at -14 C in normal air at a supersaturation level of 16%, along with measurements of R(t) and H(t). The platelike crystal is nearly flat with no sidebranching, and the ratio of prism to basal growth rates at the dendrite tips is about 30:1.

This analysis procedure assumes a fundamental tenet of the CAK model – that the basal growth rate at -14 C is independent of facet width while the prism growth rate is strongly affected by the SDAK effect at this temperature. There is no way to independently verify the accuracy of this assumption in the current experiment, as strong diffusion effects preclude a direct measurement of $\sigma_{surf}$ with any meaningful accuracy. For the crystal shown in Figure 10, a measured basal growth rate of 1.25 μm/sec gives $\sigma_{surf} \approx 1.9\%$, and the accompanying prism growth rate of 4.4 μm/sec gives $\alpha_{prism} \approx 1.0 \pm 0.1$.

Note that the CAK model assumption is giving a reasonable and not predetermined result here. It is natural to expect $\alpha_{prism} \approx 1$ on fast-growing dendrite tips, simply because of the strongly curved tip structure and the fact that molecularly rough surfaces typically exhibit $\alpha_{rough} \approx 1$ [2021Lib]. The witness-surface analysis did not force or assume $\alpha_{prism} \approx 1$, but this value was certainly expected for this fast-growing dendrite tip.

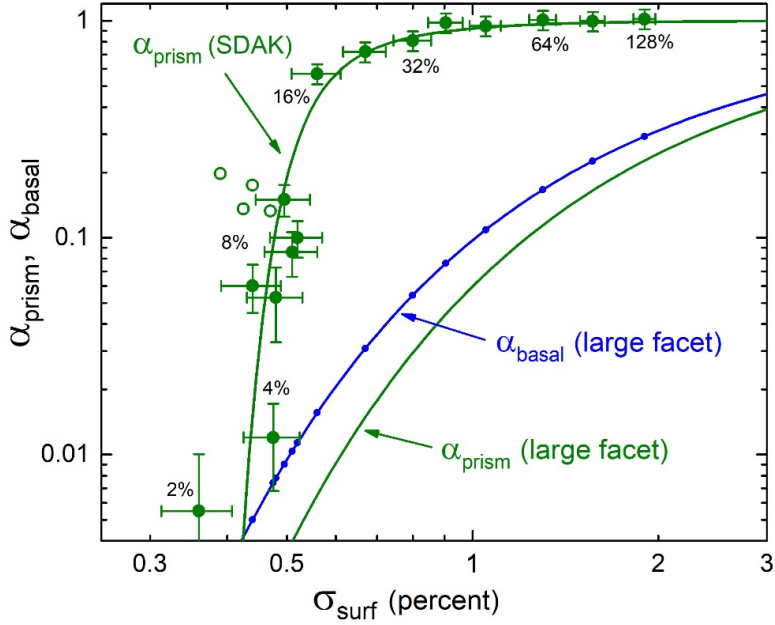

Figure 12: This plot shows $\alpha_{prism,SDAK}$ (large solid points) as a function of $\sigma_{surf}$ extracted from the growth data at -14 C presented in this paper. Numbers show values of $\sigma_{far}$ for different data points. Also shown are the CAK models (lines) for $\alpha_{basal,LF}$ and $\alpha_{prism,LF}$ on large faceted surfaces. Open points show comparable $\alpha_{prism,SDAK}(\sigma_{surf})$ data obtained from free-fall growth measurements [2008Lib1] of thin plates at -15 C, reanalyzed using the witness-surface method. Note the large increase in $\alpha_{prism}$ on narrow prism facets relative to large facets ($\alpha_{prism,SDAK} \gg \alpha_{prism,LF}$), resulting from an especially strong SDAK effect at this temperature.

Another consistency check can be obtained from the Ivantsov solution for a growing parabolic crystal [2021Lib], which gives a tip velocity of

$$v_{tip} \approx \frac{2}{B} \frac{X_0}{R_{tip}} v_{kin} \sigma_{far} \qquad (6)$$

where $B = \log(\eta_{far}/R_{tip}) \approx 10$, $\eta_{far}$ is the position of the far-away boundary (using a parabolic coordinate system with standard variables $(\xi, \eta, \varphi)$), and $\sigma_{far} = \sigma(\eta_{far})$. This gives a surface supersaturation prediction of

$$\sigma_{surf} \approx \frac{2}{\alpha B} \frac{X_0}{R_{tip}} \sigma_{far} \qquad (7)$$

and choosing $\alpha \approx 1$ and $R_{tip} \approx 2$ μm gives $\sigma_{surf} \approx 2\%$. While it is difficult to estimate $R_{tip}$ accurately from the Ivantsov parabola, the calculated $\sigma_{surf}$ is at least consistent with that extracted from the data.

Reducing the supersaturation gives similar results down to $\sigma_{far} = 16\%$, and Figure 11 shows an example at this value. The basal growth rate drops substantially when $\sigma_{far}$ changes from 256% to 16%, owing to the exponential character of $\alpha_{basal}$. The prism growth continues to be strong, however, as $\alpha_{prism}$ remains close to unity. As a result, the aspect ratio becomes more extreme at lower supersaturations, resulting in exceptionally thin, platelike crystals at 16%.

Figure 12 shows several results from these data plotted together, including the values of $\sigma_{surf}$ extracted from $v_{basal}$, the CAK model values of $\alpha_{basal}$, and the final extracted values of $\alpha_{prism,SDAK}$ from several different crystals. Going from 128% to 16%, we see that $\alpha_{basal}$ drops rapidly while $\alpha_{prism,SDAK}$ remains close to unity. In the CAK/SDAK picture, this behavior indicates a strong SDAK effect on the narrow prism facets, as the thin plate edges remain quite sharp down to $\sigma_{far} = 16\%$.

Another observation from this high-$\sigma_{far}$ region (above 16%) is that $\sigma_{surf}$ generally drops more slowly than $\sigma_{far}$, with a 2x drop in $\sigma_{far}$ giving about a 1.5x drop in $\sigma_{surf}$. This is reasonable because $\alpha_{basal}$ drops rapidly in this region, so diffusion effects tends to increase the $\sigma_{surf}/\sigma_{far}$ ratio. Again, this trend is difficult to quantify precisely, but it is roughly consistent with expectations, again lending support to our underlying assumptions of the CAK model.

The data in Figure 12 show clearly that $\alpha_{prism,SDAK}$ drop precipitously below 16%, as the prism SDAK effect diminishes rapidly at lower $\sigma_{far}$. The reason for the sudden change apparently arises from the crystal morphology, and Figure 13 illustrates this with a crystal growing at $\sigma_{far} = 11\%$. While the plate edge was exceedingly sharp at 16% (Figure 11), the edge thickness becomes substantially larger at 11% (Figure 13) Additional crystals grown at 8% look similar, except the edges become completely faceted at the lower supersaturation, yielding thick hexagonal plates.

As $\sigma_{far}$ drops to 4% and below, the e-needle growth morphology changes to a long faceted columnar structure with nearly equal basal and prism growth rates. The basal growth becomes quite difficult to measure in this regime, and the witness-surface analysis begins to lose it accuracy owing to large diffusion effects. As a result, the uncertainties in the 4% and 2% data points in Figure 12 are quite large. Nevertheless, the overall trend in the data continues, indicating a near disappearance of the SDAK effect with these crystals. At even lower supersaturations, the CAK model predicts that $\alpha_{prism,SDAK}$ would essentially equal $\alpha_{prism,LF}$ as the SDAK effect turns off.

Note in Figure 12 that $\alpha_{prism,SDAK}$ drops by two orders of magnitude while $\sigma_{surf}$ changes only from about 0.55% to 0.45%. A similar result was found in [2015Lib2] at -15 C, where direct modeling also showed a large change in prism growth rates while $\sigma_{surf}$ remained nearly constant. The CAK model explains this behavior in a natural way, as the SDAK phenomenon depends on the overall morphology of the prism faceted surface. As the prism edge narrows, the nucleation barrier drops abruptly, allowing a large increase in growth rate without a corresponding large increase in $\sigma_{surf}$. In this picture, the value of $\alpha_{prism}$ depends on temperature $T$, surface supersaturation $\sigma_{surf}$, and also quite strongly on facet width $w$. In Figure 12, it is the large change in $w$ that brings about the rapid change in $\alpha_{prism,SDAK}$ observed. This overall behavior is essentially that expected from the Edge-Sharpening Instability (ESI) that results from the SDAK effect [2021Lib, 2017Lib]. The data in Figure 12 thus strongly support the overall behavior of the CAK model, both qualitatively and quantitatively.

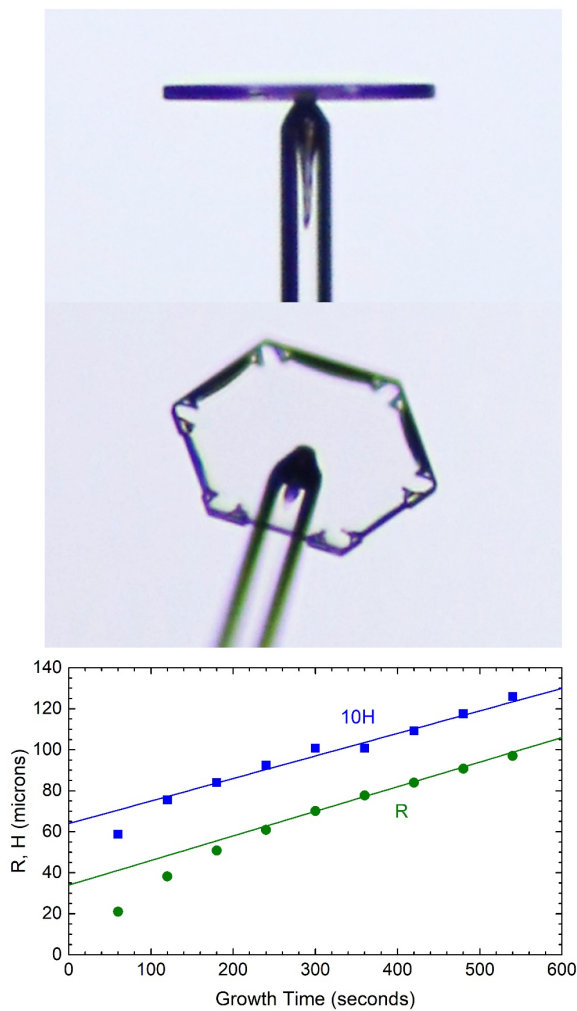

Figure 13: Two views of a snow crystal growing on the end of an e-needle at -14 C in normal air at a supersaturation level of 11%, along with measurements of R(t) and H(t). The crystal exhibits an overall thick-plate morphology, and the ratio of prism to basal growth rates to about 10:1.

## RESULTS AT -25 C

Generally blockier crystals were observed at all supersaturations measured at -25 C, with Figure 14 showing one example. The witness-surface analysis was applied as described above, yielding the results presented in Figure 15. These measurements ranged from $\sigma_{far} = 16\%$, which produced slightly flared columnar morphologies, to $\sigma_{far} = 64\%$, where the morphology was like that in Figure 14.

These results demonstrate that the SDAK effect is still present at -25 C, although it is clearly not as strong as at -14 C. The crystal morphology suggests that the one-sided

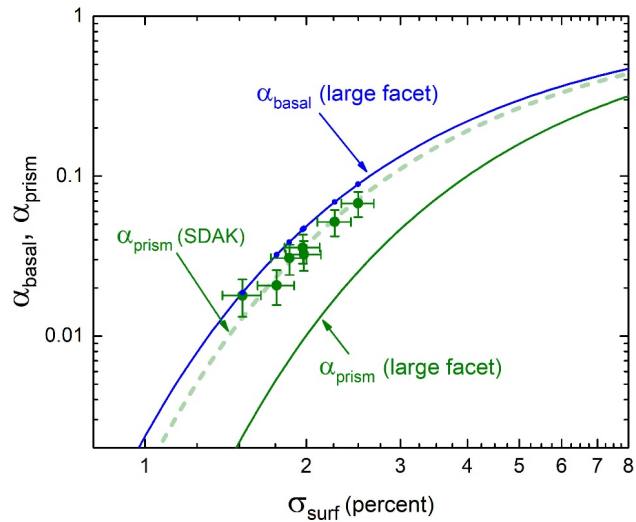

Figure 15: Data like that shown in Figure 12, but at a temperature of -25 C. The SDAK effect is weaker at this temperature, resulting in a relatively small increase in $\alpha_{prism,SDAK}$ compared to $\alpha_{prism,LF}$ on large prism facets in vacuum. Given the crystal morphology (Figure 14), this increase likely results from the one-sided SDAK effect.

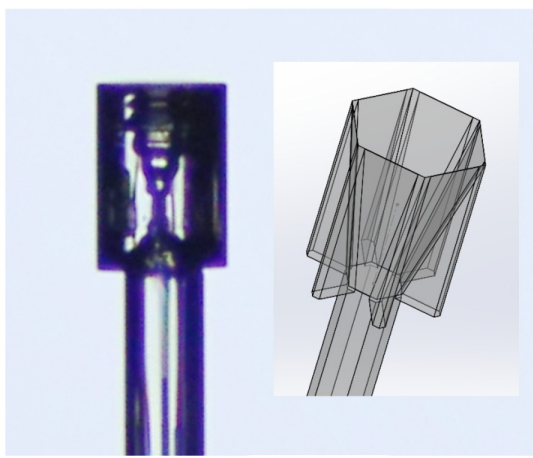

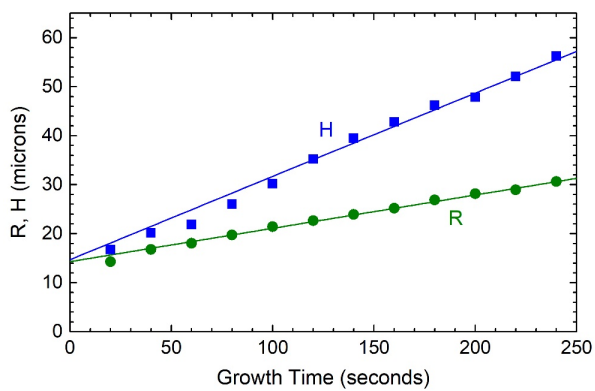

Figure 14: A blocky hollow columnar snow crystal growing on the end of an e-needle at -25 C in normal air at a supersaturation level of 45%, along with measurements of R(t) and H(t). Although difficult to see in the direct images, the morphology of this crystal appears to be roughly that of a thick conical sheath flanked by "fins", as illustrated in the inset sketch. This morphology is quite commonly found in e-needle observations. [2021Lib].

SDAK model in Figure 6 applies to all the data in Figure 15, and it appears that the Edge-Sharpening Instability is simply not strong enough at this temperature to yield the thin platelike forms seen at -14 C.

Remarkably, the measurements in Figure 15 are well described by the functional form $\alpha_{prism,SDAK}(\sigma_{surf}) \approx exp(-\sigma_{0,prism,SDAK}/\sigma_{surf})$ with $\sigma_{0,prism,SDAK} \approx 6.6\%$, while the large-facet value is $\sigma_{0,prism,LF} \approx 9.2\%$. The SDAK theory suggests that this exponential form is somewhat predicted [2019Lib1], at least in an approximate sense. The one-sided SDAK mechanism seems to produce this approximate function form at -25 C and -10 C (below), although the appearance of the stronger two-sided variation yields the more complex behavior seen in Figure 12.

The CAK model also predicts that $\alpha_{prism,SDAK}$ should reduce to $\alpha_{prism,LF}$ at $\sigma_{surf}$ values that are somewhat lower than the data points shown reported in Figure 15. In this slower growth regime, the e-needle morphology becomes essentially a solid prism, exhibiting prism facets that are nearly flat, not sloped like the example illustrated in Figure 14. Unfortunately, this morphology is accompanied by low growth rates that are difficult to measure accurately with the present e-needle experiment, along with rather large diffusion corrections that introduce substantial systematic errors into the witness-surface analysis. As a result, this additional prediction of the CAK model cannot be tested with the measurements presented here. The prediction could quite easily be tested using smaller crystals, however, so that possibility remains open for future investigations.

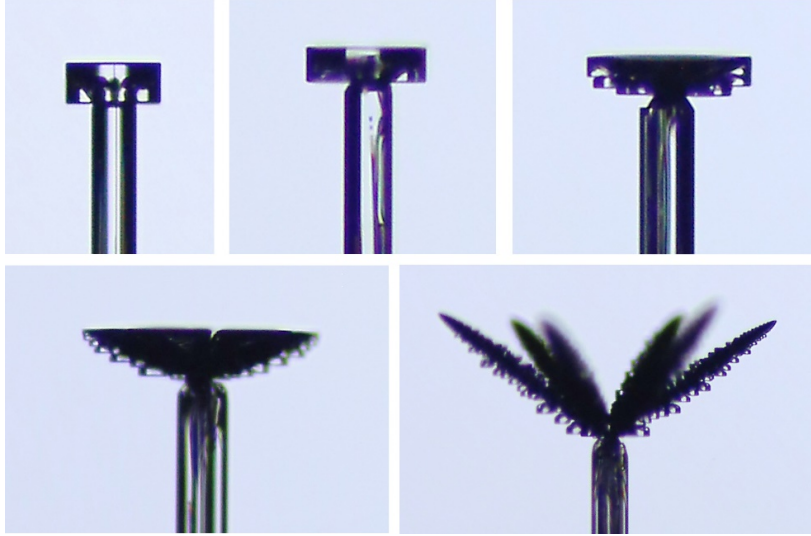

Figure 16: Ice crystals growing on e-needles in air at supersaturations of (from upper right): 8%, 16%, 32%, 64%, and 128%. At the lower values of $\sigma_{far}$, the morphologies are essentially the cone-with-fins form shown in Figure 14, but with shallower cone angles. These five images are not to scale relative to one another, and R(t) and H(t) measurements are basically like those shown at -14 C and -25 C.

## Results at -10 C

Figure 16 shows examples of similar crystals growing on e-needles at -10 C, and Figure 17 shows results using the witness-surface analysis. As with -25 C, the weaker SDAK effect yields as small reduction in the prism nucleation barrier on narrow prism facets, giving $\sigma_{0,prism,SDAK} \approx 0.85\%$ while the large-facet value is $\sigma_{0,prism,LF} \approx 1.4\%$.

Note that although blocky crystals are generally associated with snow-crystal growth at -10 C in the Nakaya diagram (Figure 1), direct observations of freely falling crystals reveal that thin plates are the norm at this temperature [2009Lib]. These free-fall data are in quite good agreement with the current results, as seen in Figure 17.

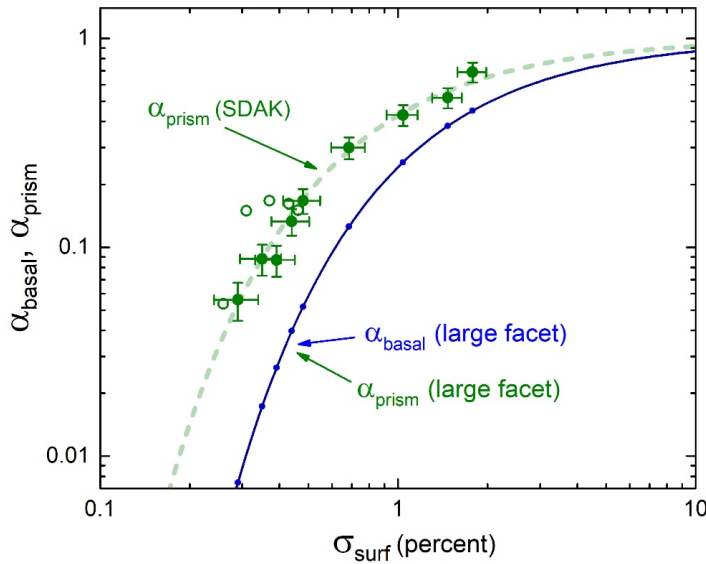

Figure 17: Additional growth data at a temperature of -10 C. The large facet curves are essentially equal at this temperature in the CAK model (as the $\sigma_{0,basal}(T)$ and $\sigma_{0,prism}(T)$ curves cross at this temperature in Figure 2), so only a single large-facet curve is shown. As at -25 C, the SDAK effect is relatively weak at -10 C, showing a relatively small increase in $\alpha_{prism,SDAK}$ relative to $\alpha_{prism,LF}$. Open points show comparable $\alpha_{prism,SDAK}(\sigma_{surf})$ data obtained from free-fall growth measurements of thin plates at -10 C [2009Lib], reanalyzed using the witness-surface method.

It is also interesting to examine the 8% crystal in Figure 16 in detail, which corresponds to the lowest-$\sigma_{surf}$ point in Figure 17. Although the prism growth is about 7x faster than the basal growth for this data point, the 8% crystal morphology appears quite blocky in overall form. A close look at the time series of images explains this, as the blocky morphology mainly results from growth of the lower basal surfaces of the block. (Determining this requires establishing a "base reference" at the lower end of the needle, which is not shown in these images.)

The overall structure of the 8% crystal is essentially that shown the sketch in Figure 14, although with a shallower cone angle. The "fins" on the crystal contact the needle as shown in the sketch, which means that there is no nucleation barrier where the fins contact the needle. Thus, the top basal surface grows quite slowly (because of its substantial nucleation barrier) while the lower edges of the fins exhibit rapid growth. The important message from this example is that morphological observations do not always tell the whole story, which requires careful measurements of growth rates on the different surfaces.

## Results at 32%

The preceding results at -10 C and -25 C show that the functional form $\alpha_{prism,SDAK}(\sigma_{surf}) \approx exp(-\sigma_{0,prism,SDAK}/\sigma_{surf})$ provides a good representation of the SDAK growth behavior, and this means that $\alpha_{prism,SDAK}(T, \sigma_{surf})$ can be reduced to a single temperature-dependent parameter $\sigma_{0,prism,SDAK}(T)$ over much of this temperature span. The measurements near the peak at -14 C do not fit the simple exponential functional form, but the fit is reasonable on either side of the peak. Over much of this temperature range, therefore, measuring $\alpha_{prism,SDAK}$ at a single value of $\sigma_{surf}$ is sufficient to estimate $\sigma_{0,prism,SDAK}$ and thus the full growth behavior of typical crystals growing in air. To explore this hypothesis, I performed a series of measurements at $\sigma_{far} = 32\%$ and converted the results to $\sigma_{0,prism,SDAK}(T)$, yielding the results shown in Figure 18. Clearly the overall shape of the $\sigma_{0,prism,SDAK}(T)$ curve is in excellent agreement with that predicted in the CAK model (Figure 4). The full data set indicates, however, that this simple picture of $\sigma_{0,prism,SDAK}(T)$ is not adequate to fully describe the growth behavior near -14 C (Figure 12), when the SDAK effect is especially strong.

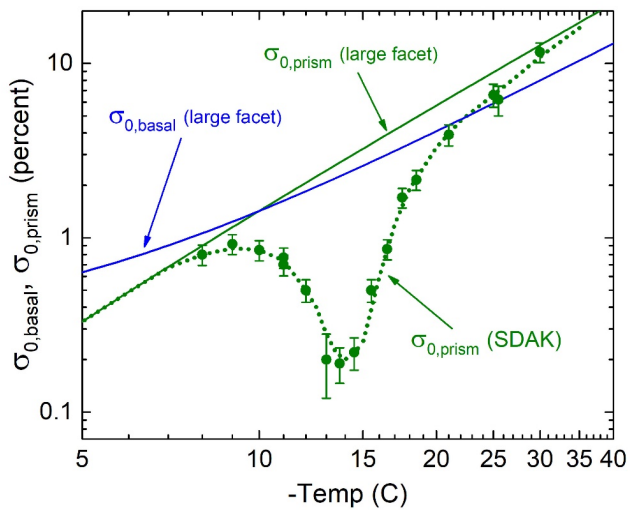

Figure 18: Measured values of $\sigma_{0,prism,SDAK}(T)$ (data points) obtained from growth measurements taken at $\sigma_{far} = 32\%$. The dotted line was drawn to guide the eye, with an obvious similarity to Figure 4 (which was a prediction from the CAK model described in [2021Lib]). Additional lines in the above figure show the large-facet CAK model (Figure 2).

## Peak Location

I also determined the local minimum point in the SDAK dip with higher accuracy by doing a series of measurements at $\sigma_{far} = 22\%$ and temperatures near -14 C. The morphology is quite sensitive to temperature at this supersaturation, with spiky branches near -14 C and hexagonal plates at both lower and higher temperatures. Measuring the radial growth as a function of temperature gives the results in Figure 19. Care was taken to determine the temperature at the crystal growth region using a calibrated thermistor with an absolute accuracy of about 0.1 C.

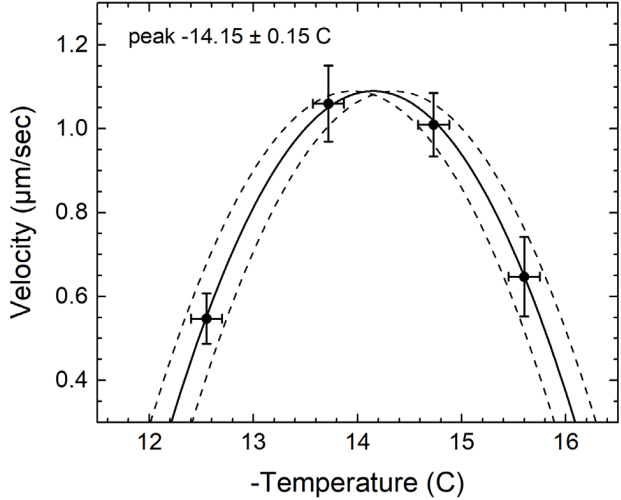

Figure 19: Measurements of the radial growth velocity of plates growing at $\sigma_{far} = 22\%$ as a function of temperature. A simple quadratic fit yields the peak location at $T = -14.15 \pm 0.15$ C. This temperature also corresponds to the minimum of the prism SDAK dip seen in Figure 18.

## ❄ Conclusions

In this paper, I used a series of ice-growth observations to test and further develop the Comprehensive Attachment Kinetics (CAK) model first presented in [2019Lib1, 2021Lib]. Being a quantitative model for calculating the attachment coefficients $\alpha_{basal}$ and $\alpha_{prism}$ over a broad range of environmental conditions, the CAK model makes a plethora of detailed predictions can be tested using targeted experimental investigations.

Because it can be so difficult to determine the surface supersaturation on a growing snow crystal, the experiments presented here do not attempt to determine $\sigma_{surf}$ directly or through diffusion modeling. Instead, I examine the relationship between $\alpha_{basal}$ and $\alpha_{prism}$ at fixed $\sigma_{surf}$, as this relationship can be investigated with meaningful precision using the witness-surface analysis. While this experimental strategy leads to somewhat model-dependent conclusions, the data nevertheless provide an important test of the overall applicability of the CAK model in snow crystal growth.

In the analysis presented above, I assumed that $\alpha_{basal}(\sigma_{surf})$ was given to good accuracy by the CAK model parameters, and then used experimental data to determine $\sigma_{surf}$ and $\alpha_{prism}(\sigma_{surf})$ from measured growth velocities. The results show good agreement with model predictions in this test, indicating $\alpha_{prism} \to 1$ at high $\sigma_{surf}$, as expected, and further revealing an SDAK dip that generally matches the CAK model behavior. Thus, the data provide strong support for the CAK model, providing a valuable endorsement of its overall picture of the physical processes governing snow crystal growth.

One important caveat is that these experimental tests are limited in their physics reach, as they do not determine $\sigma_{surf}$ in a separate, model-independent way. For this reason, we could not examine $\alpha_{basal}(\sigma_{surf})$ and $\alpha_{prism}(\sigma_{surf})$ individually, but could only examine their ratio while using $\alpha_{basal}(\sigma_{surf})$ to estimate $\sigma_{surf}$ from the model. While this analysis strategy may not be ideal, there is no better way (at present) to examine the attachment kinetics in conditions relevant to normal snow crystal growth in air, where the SDAK phenomenon is important and diffusion effects make it nearly impossible to determine $\sigma_{surf}$ with meaningful accuracy. Nevertheless, using the experimental tools I have at my disposal, the CAK model is (so far) passing all the tests I can devise [2019Lib2, 2020Lib].

The next obvious step in this experimental program is to begin at low $\sigma_{surf}$ and follow the progression of growth behaviors as $\sigma_{surf}$ is increased. At low growth rates, such that $\alpha \ll \alpha_{diff}$, diffusion effects are negligible even in air at one bar, so the resulting growth should equal the large-facet growth in the CAK model. This is true with good accuracy at -5 C [2019Lib2, 2021Lib], but it has not yet been tested thoroughly at other temperatures. Moreover, seeing the transition from large-facet growth to the SDAK curves shown in this paper would provide additional strong support for the CAK model. A crystal radius of 10 μm gives $\alpha_{diff} \approx 0.015$, so it should be possible to view this transitional behavior in air reasonably well over a broad temperature range. Examining the transition as a function of air pressure would also provide a valuable test of the model.

In this same vein, further measurements on small crystals would allow accurate determinations of $\sigma_{surf}$ using a simple monopole diffusion analysis [2021Lib], removing the need to rely on witness surfaces. In short, there are many additional targeted experiments that could explore the ice/vapor attachment kinetics in more detail, hopefully providing further confirmation of the CAK model (or suggesting its modification).

The overarching goal of this series of papers is to create a quantitative physical model of snow crystal growth. This effort has been motivated in large part by a desire to quantify and understand the Nakaya diagram, particularly the transitions between platelike and columnar growth as a function of temperature, which were first observed nearly 75 years ago. The Comprehensive Attachment Kinetics (CAK) model was devised for this purpose, and the data presented above provide strong experimental support for the model. This is the primary conclusion from these targeted experimental investigations – that the CAK model seems to be providing an adequate explanation of observed snow-crystal crystal growth behavior, both qualitatively and quantitatively.

## ❋ References